\documentclass[preprint,pre,aps]{revtex4-1}

\usepackage{graphicx}
\usepackage{amsmath}

\begin{document}

\title{ Atomic shell structure from an orbital-free-related density functional theory Pauli potential}
\author{Russell B. Thompson}
\email{thompson@uwaterloo.ca}
\affiliation{Department of Physics \& Astronomy and Waterloo Institute for Nanotechnology, University of Waterloo, 200 University Avenue West, Waterloo, Ontario, Canada N2L 3G1}
\date{July 19, 2020}

\begin{abstract}
Polymer self-consistent field theory techniques are used to find radial electron densities and total binding energies for isolated atoms. Quantum particles are modelled as Gaussian threads with ring-polymer architecture in a four dimensional thermal-space, and a Pauli potential is postulated based on classical excluded volume implemented in the thermal-space using Edwards/Flory-Huggins interactions in a mean-field approximation. Other approximations include a Fermi-Amaldi correction for electron-electron self-interactions, a spherical averaging approximation to reduce the dimensionality of the problem, and the neglect of correlations. Polymer scaling theory is used to show that the excluded volume form of Pauli potential reduces to the known Thomas-Fermi energy density in the uniform limit.  Self-consistent equations are solved using a bilinear Fourier expansion, with radial basis functions, for the first eighteen elements of the periodic table. Radial electron densities show correct shell structure, and the errors on the total binding energies compared to known binding energies are less than $9\%$ for the lightest elements and drop to $3\%$ or less for atoms heavier than nitrogen. More generally, it is suggested that only two postulates are needed within classical statistical mechanics to achieve equivalency of predictions with static, non-relativistic quantum mechanics: First, quantum particles are modelled as Gaussian threads in four dimensional thermal-space and, second, pairs of threads (allowing for spin) are subject to classical excluded volume in the thermal-space. It is shown that these two postulates in thermal-space become the same as the Heisenberg uncertainty principle and the Pauli exclusion principle in three dimensional space. 
\end{abstract}

\maketitle

\section{Introduction}

Wolfgang Pauli introduced the exclusion principle in 1925 to explain certain observed properties of atoms. According to Pauli, the principle states that, in an atom, two or more electrons cannot have the same values for all four quantum numbers \cite{Pauli1925}. While well defined mathematically, the exclusion principle is an unintuitive quantum mechanical concept dependent on wave-particle duality. It causes regions of space, orbitals in atoms for example, to be ``occupied'' and prevents the presence of other quantum particles in those regions. On one hand, if one views quantum particles as true \emph{particles}, there is no reason why a localized entity like a point-particle should prevent other particles from existing in an extended region. On the other hand, if quantum particles are viewed as waves, the non-local exclusion makes sense, but then one is confronted by the measurement problem, that is, explaining the nature of wave function collapse to a particle upon measurement.

Before quantum mechanics, the closest classical analogue of the exclusion principle, excluded volume, was postulated as an intuitive, common sense principle -- two objects cannot fill the same space at the same time \cite{Margenau1944}. This has been supplanted by the exclusion principle, which nowadays gives the origin of excluded volume. The Pauli principle is therefore a postulate of non-relativistic quantum mechanics, and although it is not always explicitly enumerated as a postulate in textbooks, it cannot be proven through ideas of indistinguishability as is commonly suggested \cite{Kaplan2013}. 

The Pauli principle is very important for orbital-free versions of density functional theory (DFT). The theorems of DFT allow the use of spatially dependent single particle number densities $n({\bf r})$ instead of wave functions to obtain predictions from quantum mechanics through a ground-state energy expression that is a functional of the density \cite{Hohenberg1964, vonBarth2004, Becke2014, Jones2015}. Time dependent, temperature dependent, and relativistic versions of DFT also exist \cite{vonBarth2004, Mermin1965}. In DFT, correlations and exchange effects are typically grouped together in the ``exchange-correlation'' functional $E_{xc}[n]$ which, if known exactly, permits exact quantum mechanical solutions. In reality, $E_{xc}[n]$ must be approximated, but DFT is nonetheless one of the most powerful modern techniques for solving problems in quantum chemistry and physics \cite{Hohenberg1964, vonBarth2004, Becke2014, Jones2015}. The most common method of solving DFT is through the use of ``orbitals'', that is, a set of solutions of an eigenvalue equation which indirectly allows one to find the density through a sum over the occupied orbitals \cite{Kohn1965}. This ``Kohn-Sham'' DFT (KS-DFT) incorporates the Pauli exclusion principle ``by hand'' through the sum over orbitals, leaving the exchange-correlation functional to estimate the energy attributable to the Pauli principle. A computationally more efficient alternative to KS-DFT is orbital-free DFT (OF-DFT) which, as the name implies, avoids the use of orbitals and computes the structure and energy of quantum systems directly using only the density \cite{Carter2018, Wang2000, March2010, Karasiev2014}. The disadvantage of OF-DFT is that in addition to approximating the exchange-correlation functional, the non-interacting kinetic energy is not known exactly as it is in KS-DFT. It has been shown that correctly approximating the kinetic energy is equivalent to incorporating the Pauli exclusion principle \cite{Finzel2015b, Finzel2016b, Finzel2017}.

Recently, an alternative derivation of OF-DFT was presented that did not use the theorems of DFT, but instead used quantum statistical mechanics to derive a temperature-dependent free energy functional of the quantum particle density \cite{Thompson2019}. The derivation used techniques from polymer self-consistent field theory (SCFT) \cite{Matsen2020, Matsen2006, Qiu2006, Fredrickson2002, Schmid1998} and reduced to KS-DFT, assuming a strict enforcement of the Pauli exclusion principle. Without this assumption, the formalism is bosonic, like other OF-DFTs, so an exchange-correlation term including the Pauli exclusion principle was needed to make predictions for electronic systems. A rigorous but simple test of exchange-correlation functionals, including different approximations of the Pauli principle, is the atomic system. Atoms have very inhomogeneous electron densities, showing shell-structure due to the Pauli exclusion principle. In reference \onlinecite{Thompson2019}, correlations were ignored, and a simple shell-structure-based (SSB) Pauli potential together with a local density approximation exchange term (LDAX), as presented by Finzel \cite{Finzel2015}, was adopted. The results showed correct qualitative shell-structure, essentially identical to that found by Finzel \cite{Finzel2015}, but the energies of the structures were far from literature values \cite{NIST}. Also, the SSB Pauli potential is artificially incorporated as an external potential, as it is a series of step functions that are calibrated independent of the electron density $n({\bf r})$. It may be awkward to modify the spherically symmetric step functions for situations other than atoms, such as molecules or solid state materials.

\begin{figure}
\includegraphics[width=0.5\textwidth]{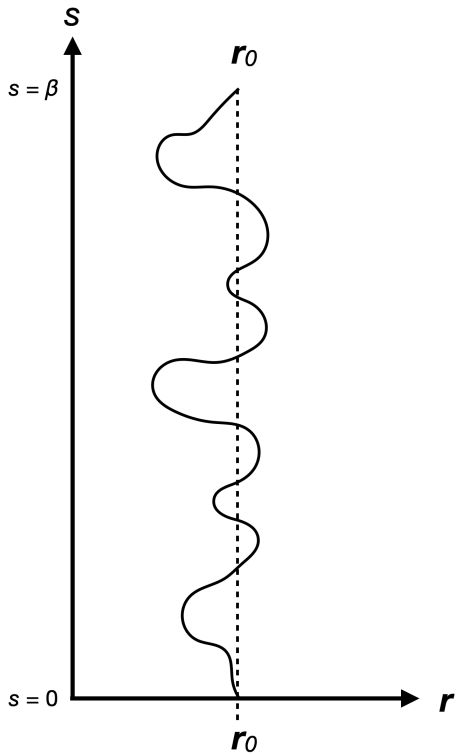}
\caption{Schematic of a quantum particle contour. The abscissa represents 3D space and the ordinate is the thermal variable, $s = 1/k_BT$. Note that the two ends of the contour are at the same spatial location, ${\bf r}_0$, giving it a ``ring'' architecture.}
\label{fig:ring}
\end{figure}

It was found in reference \onlinecite{Thompson2019} that the SCFT equations resulting from the quantum statistical mechanics derivation are identical in form to the SCFT equations of a system of ring polymers derived from classical statistical mechanics \cite{Kim2012}. Ring polymers, which are macromolecules that formed closed rings without free ends, are mathematically parameterized in SCFT along their backbone by a contour variable embedded in three dimensional space, whereas the contour parameter for quantum particles corresponds to an independent variable, specifically, an inverse thermal energy, $\beta = 1/k_BT$, where $k_B$ is Boltzmann's constant and $T$ is temperature, as shown in figure \ref{fig:ring}.
The temperature dimension behaves mathematically as an imaginary time variable. This isomorphism between quantum statistical mechanics and imaginary time ring polymer classical statistical mechanics is not new \cite{Chandler1981, Ceperley1995}. It emerges from Feynman's path integral methodology \cite{Feynman1953a,  Feynman1953b, Feynman1953c, Feynman1965}, as discussed by Ceperley \cite{Ceperley1995} in the context of bosons, and it forms the basis of path integral Monte Carlo and path integral molecular dynamics simulations.

In polymer physics, in addition to polymer Monte Carlo and molecular dynamics, one of the most important theoretical tools is SCFT.  For quantum systems however, classical SCFT is, at first glance, not applicable. As mentioned, reference \onlinecite{Thompson2019} showed that SCFT reduces to DFT assuming a strict enforcement of the Pauli exclusion principle. This demonstrates that DFT fills the role in quantum systems that SCFT holds in polymer systems. In other words,  DFT is the quantum isomorph of classical SCFT, where quantum particles are represented as ring polymers in a four dimensional imaginary time thermal-space, as suggested by Chandler and Wolynes \cite{Chandler1981} and Ceperley \cite{Ceperley1995}. 

As mentioned, the derivation of DFT in reference \onlinecite{Thompson2019} did not use the theorems of DFT. The DFT theorems show that the properties of a many-body wave function can be described by a single particle electron density \cite{Jones2015}, but this one-to-one mapping can be reversed. This leads to the possibility of an interpretation of quantum mechanics based on the Feynman quantum-classical isomorphism, according to the following reasoning:
\begin{enumerate}
\item Postulate 4D thermal-space polymers within classical statistical mechanics.
\item Following reference \onlinecite{Thompson2019}, derive ring polymer SCFT, which is equivalent to quantum DFT assuming the enforcement of the Pauli exclusion principle. 
\item Quantum DFT is in turn equivalent to quantum mechanics, from the theorems of DFT \cite{Hohenberg1964, vonBarth2004, Becke2014, Jones2015}.
\end{enumerate}
Ideally, all the predictions of quantum mechanics should be obtainable from classical statistical mechanics with an extra temperature (imaginary time) dimension. In reality, several issues have been ignored. In particular, step 2 assumes the Pauli principle is rigorously enforced, which is not the case for the orbital-free-related DFT derived in \onlinecite{Thompson2019}. Nonetheless, in reference \onlinecite{Thompson2019}, it was speculated that the Feynman quantum-classical isomorphism could be used to interpret quantum mechanics in an ensemble picture as the classical statistical mechanics of polymer-like objects in a four dimensional thermal-space. From this perspective, the SCFT equations can be interpreted as a ``hidden-variables'' theory of quantum mechanics, with the hidden variables being conformational degrees of freedom of the non-local polymer-like contours that represent quantum particles. That is, intra-particle thermal correlations are the hidden degrees of freedom.

It is the purpose of this paper to explore this possibility further with a focus on incorporating the Pauli exclusion principle into the Feynman isomorphism. A simple expression for the Pauli potential will be given that gives proper shell structure for atoms and quantitative atomic binding energies, and that is also versatile enough to apply to other systems such as molecules or solid state materials, although these applications will not be explored in this work. The Pauli potential is derived in section 2 through the classical 4D thermal-space picture, replacing the Pauli exclusion principle with the older classical excluded volume concept. It is postulated that the thermal contours that mathematically represent quantum particles cannot occupy the same space as each other, with the exception of one other contour each to allow for spin. The basic polymer excluded volume formalism of Edwards \cite{Edwards1965}, adapted for inter-polymer instead of intra-polymer interactions, is used, which is essentially the same as the Flory-Huggins interactions often used between polymers of different chemical species \cite{Matsen2020, Matsen2006, Fredrickson2002, Schmid1998, deGennes1979}. In the mean-field approximation, this Pauli potential, together with a crude Fermi-Amaldi self-interaction correction to the mean-field electron-electron Coulomb expression \cite{Ayers2005}, and an ionic Coulomb potential, are the only interactions used in this paper to study the atoms hydrogen through argon in section 3. Correlations are ignored, but despite this, the electron densities show correct shell-structure, and the total binding energies are quantitatively correct, although not surprisingly, they are not close to chemical accuracy. To further check the 4D thermal-space excluded volume model of the Pauli potential, polymer scaling theory, following the ideas of de Gennes \cite{deGennes1979}, is used to verify that in the uniform limit with high numbers of electrons, the Thomas-Fermi 5/3 power law dependency of the electron density is found. In section 4, the idea of quantum particles being non-local polymer-like contours in a 4D thermal-space, and the idea of these contours possessing classical excluded volume in the 4D space, are taken together as two postulates, within classical statistical mechanics, necessary to reproduce the predictions of static, non-relativistic quantum mechanics. Conclusions are drawn in section 5.

\section{Theory}

Feynman's quantum-classical mapping means that the free energy $F$ of a system of $N$ quantum particles subject to an external potential $U_{\rm ext}$ can be derived from quantum statistical mechanics, as in appendix A of reference \onlinecite{Thompson2019}, or from the classical statistical mechanics of ring polymers, as in reference \cite{Kim2012}. Either way, the result is  
\begin{equation}
F[n,w] = -\frac{N}{\beta}\ln Q(\beta) - \int d{\bf r} w({\bf r}) n({\bf r},\beta) + U[n]    \label{FE3}
\end{equation}
where $n({\bf r},\beta)$ is the quantum particle density as a function of position ${\bf r}$ and inverse temperature $\beta = 1/k_BT$. $Q(\beta)$ is a single particle partition function subject to the field $w({\bf r})$. The SCFT method of derivation for equation (\ref{FE3}) is a rigorous first principle path integral approach that has been extensively reviewed for a variety of polymer systems \cite{Matsen2020, Matsen2006, Qiu2006, Fredrickson2002, Schmid1998} and so will not be repeated here. Physically, the first two terms on the right hand side of equation (\ref{FE3}) give a combination of configurational and translational entropy of a polymer subject to the field $w({\bf r})$. The polymer configurational entropy is itself isomorphic to the quantum kinetic energy. The expression for the quantum kinetic energy in (\ref{FE3}) is exact to the extent that the fields $w({\bf r})$ are correct; the fields are generated by the third term on the right hand side of (\ref{FE3}), which is
\begin{equation}
U[n] = U_{\rm int}[n] + U_{\rm ext}[n]     \label{U1}
\end{equation}
where $U_{\rm int}$ is the sum of internal potentials between quantum particles, such as electron-electron interactions. The external potential $U_{\rm ext}$ is due to, for example, the ionic Coulomb potential in atomic systems. In appendix A of reference \onlinecite{Thompson2019}, the functional (\ref{FE3}) is extremized to produce a set of equations that is solved numerically and self-consistently in order to find the quantum particle density $n({\bf r},\beta)$. The reader is referred to reference \cite{Thompson2019} for the full details. As mentioned, these equations are identical to equations derived through \emph{classical} statistical mechanics for ring architecture polymers using SCFT \cite{Kim2012}, except that the parameter $s$ for polymers is a spatial parametrization of the polymer contour embedded in real space, whereas for quantum particles it is an independent variable describing a ``thermal trajectory'' running from zero to $\beta = 1/k_BT$. The speculation of reference \onlinecite{Thompson2019} is that non-relativistic quantum mechanics can be derived using classical statistical mechanics through SCFT on an ensemble of polymer-like objects which represent quantum particles in a four dimensional thermal-space consisting of ${\bf r}$ and $\beta$. 

In order to test and extend this hypothesis, one can attempt to describe the Pauli exclusion principle within the SCFT framework using only classical concepts. The appropriate classical analogue to the exclusion principle is excluded volume. In polymer SCFT, excluded volume is often included using the Edwards expression, which implements a Dirac delta function energy penalty for polymer overlaps \cite{Edwards1965}. Normally, the Edwards potential is applied to a single polymer chain interacting with itself, but in the Pauli context, it should be applied between different polymer-like quantum particles. This is mathematically the same as the polymer Flory-Huggins interaction between distinct chemical species, which also uses a Dirac delta function energy penalty. In this work, an ensemble of each individual quantum particle will be treated as a distinct species, or allowing for spin, each pair of particles will be its own species and will interact with other pairs through a Flory-Huggins-type potential. The term ``pairs'' will mean each ``species'' can have up to two particles. The sum of the electron densities of all of these species will give the total electron density; other potential terms of (\ref{U1}), for example the electron-electron Coulomb potential and the external potential, will continue to be functionals of the total electron density. 

\begin{figure}
\includegraphics[width=1.0\textwidth]{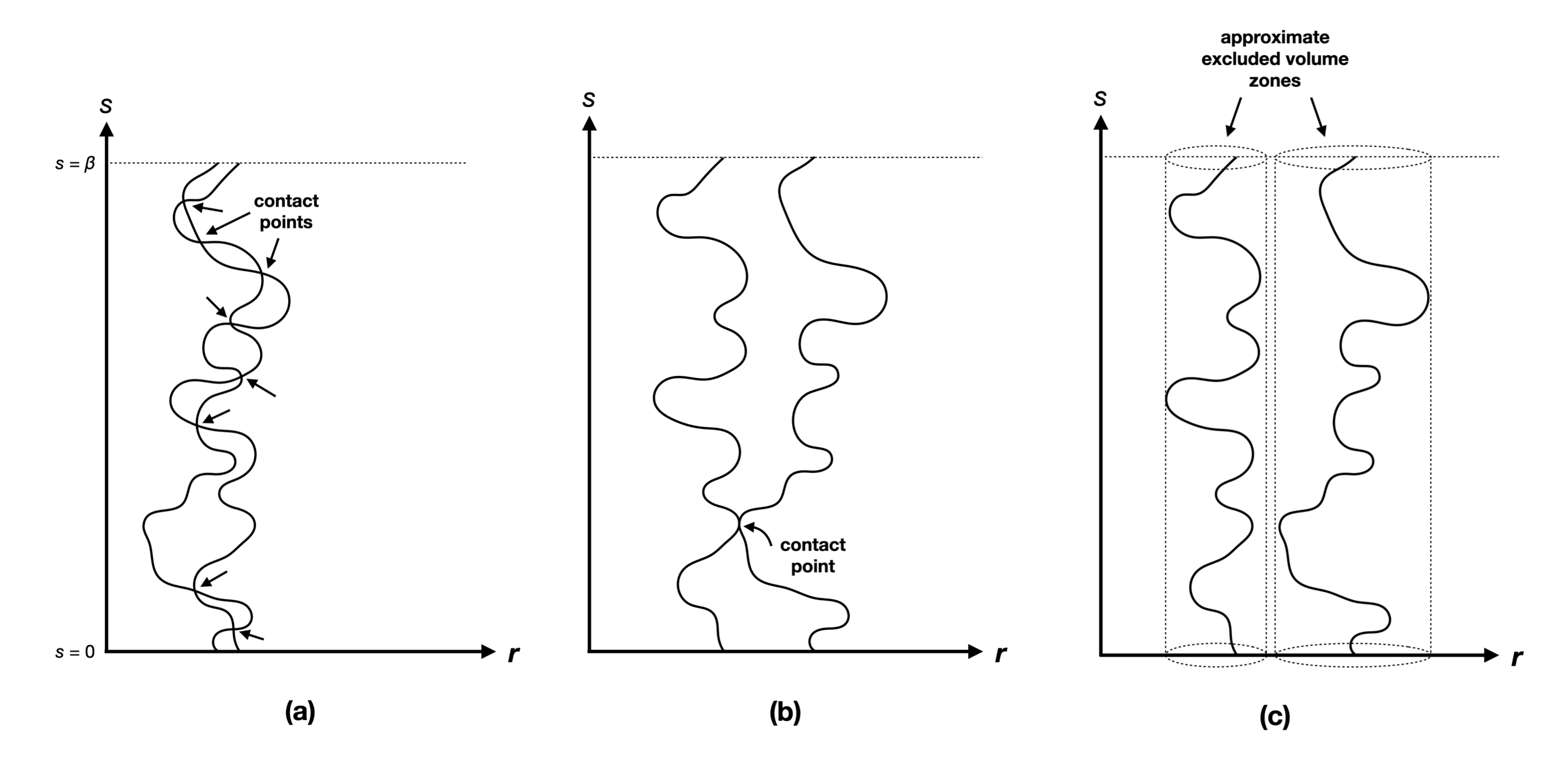}
\caption{Schematic of two quantum particle contours in 4D with the abscissas representing 3D space and the ordinates the thermal variable, $s = 1/k_BT$. (a) Two contours very close together showing many points of excluded volume contact, indicated by the arrows. (b) Two contours somewhat separated with fewer overlapping contact points. (c) Two contours further apart again, free of overlaps. This is approximated in the model by excluded volume interactions projected out into 3D, shown schematically by the cylindrical regions.}
\label{fig:Vex}
\end{figure}

Following the Flory-Huggins formalism, the Pauli energy between quantum particle pairs should be an internal potential of the form
\begin{equation}
U_P[\{n\}] = \frac{1}{2} \sum_{\substack{ij \\ i \neq j}} \int\int d{\bf r} d{\bf r}^\prime n_i({\bf r},\beta) V_{\rm xv}(|{\bf r}-{\bf r}^\prime|) n_j({\bf r}^\prime,\beta)  \label{UP1}
\end{equation}
assuming a mean field expression, where $\{n\}$ is the set of all quantum particle pair densities, the summations are over all pairs, and $V_{\rm xv}(|{\bf r}-{\bf r}^\prime|)$ is an excluded volume interaction energy of the form 
\begin{equation}
V_{\rm xv}(|{\bf r}-{\bf r}^\prime|) = g^{-1}_0 \delta({\bf r}-{\bf r}^\prime)  . \label{Vex1}
\end{equation}
The magnitude of the constant prefactor $g^{-1}_0$ needs to be specified, but it will have units of an inverse density of states. Expression (\ref{Vex1}) is approximate, since rigorously it should be of the form
\begin{equation}
V_{\rm xv}(|{\bf r}(s)-{\bf r}^\prime (s)|) = g^{-1}_0 \delta({\bf r}(s)-{\bf r} ^\prime (s))  \label{Vex2}
\end{equation}
where ${\bf r}(s)$ is a parametrized curve describing the contour of the polymer-like quantum particle in the 4D thermal-space. Equation (\ref{Vex2}) indicates an excluded volume energy penalty only when two contours are at the same place \emph{for the same value of the contour}. This is analogous to particles having to be at the same place \emph{at the same time} in order to feel excluded volume. Ceperley has discussed how interactions should occur only between different contours and only at the same imaginary time-slice \cite{Ceperley1995}. For regular polymers, the contours are embedded in space, so any segment can bump into any other segment, and equation (\ref{Vex1}) is used. For quantum particles, the parameter $s$ is \emph{not} embedded in 3D, but is an independent time-like variable, so one should use (\ref{Vex2}), but this is difficult to implement. Using (\ref{Vex1}) instead,  which ``projects out'' the 4D contour into 3D space, gives a good approximation, although it will over-estimate excluded volume somewhat. The situations described by (\ref{Vex1}) and (\ref{Vex2}) are shown in figure \ref{fig:Vex}. 
Using (\ref{Vex1}) neglects inter-contour correlations and so it is a mean field approximation. Putting (\ref{Vex1}) into (\ref{UP1}) gives
\begin{equation}
U_P[\{n\}] = \frac{1}{2} \sum_{ij} g^{-1}_{ij} \int n_i({\bf r},\beta) n_j({\bf r},\beta) d{\bf r}   \label{UP2}
\end{equation}
where
\begin{equation}
g^{-1}_{ij} \equiv (1-\delta_{ij})g^{-1}_0  . \label{g0inv}
\end{equation}
Since the excluded volume effect between 4D quantum particle contours should be universal, the parameter $g^{-1}_0$ should not be free. It can be calibrated by comparing predictions to any experimental result; in this way, no assumption needs to be made about the internal structure of an electron contour, since the internal structure of fundamental particles are not known. For simplicity, $g^{-1}_0$ will be set here by comparing the Pauli energy to a theoretical ``jellium'' state, that is, a uniform gas of quantum particles with no interactions other than the exclusion principle. The energy density of this system for electrons is the Thomas-Fermi energy density \cite{Carter2018}, derived by Ashcroft and Mermin in the Sommerfeld model to be \cite{Ashcroft1976}
\begin{equation}
\frac{U_P}{V} = c_0 n_0^{5/3}  \label{TF}
\end{equation}
where $n_0=N/V$ is the average electron density, $V$ is the volume, and
\begin{equation}
c_0 = \left(\frac{3}{10}\right) (3 \pi^2)^{2/3} .  \label{c0}
\end{equation}
The Thomas-Fermi expression (\ref{TF}) is valid for uniform densities with large numbers of electrons, so that the Fermi surface can be approximated as a continuous sphere. Under these conditions, as shown in the appendix, the Pauli energy expression (\ref{UP2}) can be phrased as an energy density and becomes
\begin{equation}
\frac{U_P}{V} = \frac{g^{-1}_0}{2}n_0^2  .  \label{TF2}
\end{equation}
As expected, (\ref{TF2}) overestimates the strength of the Pauli interaction by a factor of $n_0^{1/3}$. Ignoring this weak factor, one gets an estimate for $g^{-1}_0$ as
\begin{equation}
g^{-1}_0 = 2c_0  \approx 5.742468   \label{g0inv2}
\end{equation}
leaving no free parameters in the theory. The missing factor $n_0^{1/3}$ could in principle be accounting for by correlation correction terms to the energy, but the atomic systems that will be the focus of this study do not necessarily have high $N$ and are certainly not homogeneous. Thus correlation corrections based on Thomas-Fermi are not within the scope of this study. Rather, it is beneficial to scrutinize the correlation-free estimate based on (\ref{g0inv2}) to see the size of the errors that are produced. For the uniform, large electron number case appropriate to the Thomas-Fermi expression (\ref{TF}), one can use polymer scaling theory to avoid neglecting correlations. In polymer systems, the mean field description systematically also overestimates excluded volume, in contrast to scaling theory \cite{deGennes1979}. For the present electron-contour case, it is shown in the appendix, by adapting the arguments of de Gennes \cite{deGennes1979}, that the model of polymer-like excluded volume interactions gives an energy density dependence on the electron density of
\begin{equation}
\frac{U_P}{V} \sim n_0^{5/3}  \label{TF3}
\end{equation}
consistent with Thomas-Fermi.

The canonical ensemble free energy functional (\ref{FE3}) can be generalized to include densities for pairs of quantum particles as
\begin{equation}
F[\{n\},\{w\}] = -\frac{1}{\beta}\sum_i N_i \ln Q_i(\beta) - \sum_i \int d{\bf r} w_i({\bf r}) n_i({\bf r},\beta)  + U[\{n\},n]     \label{FE4}
\end{equation}
where $N_i$ are the numbers of quantum particles in each pair (0, 1 or 2 particles in each ``pair''), $n_i({\bf r},\beta)$,  $w_i({\bf r})$ and $Q_i(\beta)$ are the densities, fields and single particle partition functions, respectively, of each pair and $\{n\}$ and $\{w\}$ are the sets of all pair densities and fields, respectively. The summations are over all pairs. The total number of quantum particles and the total density, respectively, are 
\begin{eqnarray}
N &=& \sum_i N_i  \label{Ntot}  \\
n({\bf r},\beta) &=& \sum_i n_i({\bf r},\beta)  . \label{ntot}
\end{eqnarray}
The total potential $U[\{n\},n]$ is
\begin{equation}
U[\{n\},n] = U_0[n] + \sum_i U_P[\{n\}]  \label{Utot}
\end{equation}
where $U_P[\{n\}]$ is given by (\ref{UP2}) and $U_0[n]$ accounts for all remaining internal and external potentials that depend only on the total density. In most cases, the external potential will be a Coulomb potential due to ions. The remaining internal potential, for electronic systems, will be the electron-electron Coulomb interaction. Expressions for both of these are given in appendix E of reference \onlinecite{Thompson2019}. Normally, self-interaction corrections to the Hartree electron-electron Coulomb interaction are included in the exchange-correlation functional of DFT. Here however, the Hartree term is modified using the Fermi-Amaldi pre-factor of $(N-1)/N$ \cite{Ayers2005} so that it will exactly account for self-interactions for $N=1$ and asymptotically large $N$. For most values of $N$, the Fermi-Amaldi self-interaction correction is considered to be crude \cite{Perdew1981}, however it is an orbital-free expression that is very simple to implement, and, as shall be shown in the Results section, it gives reasonable results. Since no correlations are included, the entirety of exchange-correlation effects in this work are contained in the Fermi-Amaldi and Pauli terms.

Following reference \onlinecite{Thompson2019}, the free energy functional (\ref{FE4}) can be varied with respect to all pair densities and fields to give the set of self-consistent equations
\begin{eqnarray}
w_i({\bf r}) &=& \frac{\delta U[\{n\},n]}{\delta n_i({\bf r},\beta)}  \label{w4} \\
n_i({\bf r},\beta) &=& \frac{n_{0i}}{Q_i(\beta)} q_i({\bf r},{\bf r},\beta)    \label{n4}
\end{eqnarray}
and
\begin{equation}
Q_i(\beta) = \frac{1}{V} \int d{\bf r} q_i({\bf r},{\bf r},\beta)   \label{Q4}
\end{equation}
with
\begin{equation}
\frac{\partial q_i({\bf r}_0,{\bf r},s)}{\partial s} = \frac{\hbar^2}{2m} \nabla^2 q_i({\bf r}_0,{\bf r},s) - w_i({\bf r}) q_i({\bf r}_0,{\bf r},s)  \label{diff4}
\end{equation}
subject to the initial conditions
\begin{equation}
q_i({\bf r}_0,{\bf r},0) = V \delta({\bf r}-{\bf r}_0)    \label{init4}
\end{equation}
where $n_{0i} \equiv N_i/V$. The functional derivatives in equations (\ref{w4}) can be carried out on (\ref{Utot}) to give the same total density dependent fields as in appendix E of reference \onlinecite{Thompson2019}, but with the extra Fermi-Amaldi pre-factor on the electron-electron potentials $w_{{\rm ee},i}({\bf r})$. Each total field $w_i({\bf r})$ will be distinct due to the functional derivatives of the Pauli terms (\ref{UP2}) which give contibutions
\begin{equation}
w_{P,i}({\bf r},\beta) = g^{-1}_0 \sum_{j \neq i} n_j({\bf r},\beta) .  \label{wP}
\end{equation}
Equations (\ref{w4}) - (\ref{wP}) are solved numerically and self-consistently, with the computationally limiting factor being the solution of the diffusion equations (\ref{diff4}). Matsen has suggested a bilinear Fourier series expansion method \cite{Matsen2019}, used in reference \onlinecite{Thompson2019}, that greatly reduces the computational cost of solving ring polymer systems, and this method is used again here. Equations (\ref{w4}) - (\ref{wP}) are expanded in terms of the Fourier basis functions that have the symmetry of the system of interest, and the spectral equations are solved for the Fourier coefficients. Atomic systems are studied in this paper since they are a simple and yet non-trivial application, due to their severely inhomogeneous electron densities. The spherical symmetry of the ensemble average electron densities in atomic systems allows the choice of spectral basis set to be zeroth order spherical Bessel functions, which is a complete orthonormal set -- see reference \cite{Thompson2019} appendix D \footnote{Equation (D12) of reference \onlinecite{Thompson2019} should not have a summation.}. Equations (\ref{w4}) - (\ref{wP}) expanded in spherical Bessel functions give the same expressions as in reference \onlinecite{Thompson2019} appendix E, but now with the inclusion of the Fermi-Amaldi pre-factor and the addition of the Pauli potential. Equation (\ref{wP}) written in spectral form is
\begin{equation}
w_{P,i}^k = g^{-1}_0 \sum_{j \neq i} n_j^k  \label{wPspectral}
\end{equation}
where $n_j^k$ and $w_{P,i}^k$ are the Fourier coefficients for the densities and Pauli fields, respectively. The SCFT system of equations (\ref{w4}) - (\ref{wPspectral}) is solved for the first eighteen atoms of the periodic table, using the same numerical approach as in reference \onlinecite{Thompson2019} (where full numerical details can be found), until the field coefficients stop changing by less than one part in $10^{-9}$ according to the square of an L2-norm.

\section{Results}

The free energies of equation (\ref{FE4}) were calculated for isolated atoms of hydrogen through argon using values of $\beta$ that approach zero temperature. Specifically, a value of $\beta=80$ was large enough to approach zero temperature, and for some cases, $\beta=20$ or less was sufficient. Results are shown in Hartree atomic units in table \ref{tab:elemFE}, which also gives NIST values for the binding energies \cite{NIST}. 
\begin{table}
\centering
\caption{SCFT calculated binding energies for the first 18 elements of the periodic table compared with NIST values \cite{NIST}. Energies are in Hartree atomic units.}
\begin{tabular}{cccc}
\hline \hline
Element & \parbox{4.2cm}{SCFT Binding Energy} & \parbox{4cm}{NIST Binding Energy} & \% Difference \\
\hline 
H &  0.5000000 & 0.4997332 & 0.05 \\
He & 2.7616721 & 2.9033858 & 4.9 \\
Li & 6.85169 & 7.4779785 & 8.4 \\
Be & 13.4755 & 14.668442 & 8.1 \\
B & 23.010 & 24.658095 & 6.7 \\
C & 35.812 & 37.855785 & 5.4 \\
N & 52.225 & 54.611615 & 4.4 \\
O & 72.60 & 75.10984 & 3.3 \\
F & 97.26 & 99.8071 & 2.6 \\
Ne & 126.57 & 129.05245 & 1.9 \\
Na & 159.1 & 162.432 & 2.1 \\
Mg & 195.8 & 200.323 & 2.3 \\
Al & 237.0 & 242.7275 & 2.4 \\
Si & 282.6 & 289.898 & 2.5 \\
P & 332.9 & 341.98 & 2.7 \\
S & 387.8 & 399.085 & 2.8 \\
Cl & 447.9 & 461.44 & 2.9 \\
Ar & 512.8 & 529.22 & 3.1 \\
\hline
\end{tabular}
\label{tab:elemFE}
\end{table}
Numerical errors on the SCFT results are estimated to be $ \lesssim 0.2\%$ for all elements, and $ \ll 0.1 \%$ for most. 

The percent deviations between the SCFT binding energies and the NIST values given in table \ref{tab:elemFE} show that, although far from chemical accuracy, the energies are very good. The percent deviation is highest for lithium, which is not surprising since the Fermi-Amaldi self-interaction correction is poor there, but improves as $N$ increases, as expected. The  deviation for elements heavier than nitrogen is roughly $3\%$ or below.

In addition to the Fermi-Amaldi approximation, and the neglect of correlations, there is also a spherical averaging approximation which affects atoms heavier than beryllium. In principle, a maximum of 2 electrons should be allowed for each pair in a fully three dimensional SCFT calculation, and atomic states such as 2p or 3p would spontaneously appear, breaking spherical symmetry, just as low symmetry morphologies arise spontaneously in numerical polymer SCFT results \cite{Fredrickson2002, Rasmussen2002a, Rasmussen2002b}. For computational efficiency however, known atomic shell structure is input to reduce the numerical burden, so that non-spherical sub-shells are lumped together with spherical shells. For example, three diffusion equations should be solved for boron $1s^2 2s^2 2p^1$. Instead only two are solved, one for the pair of $1s^2$ electrons and another for the three $2s^2 2p^1$ electrons. Since the energy level difference between the $2s$ and $2p$ states is small, this approximation is expected to be reasonable. Indeed, the errors in table \ref{tab:elemFE} show no significant consequence for this approximation. In principle, one could use an effective  $g^{-1}_0$ parameter for states containing more than two electrons in the spirit of a pseudo-potential, but this has not been done here. Rather, it is desirable to see the full effect of all approximations. Nonetheless, the SCFT formalism could possibly be made to go smoothly from an all-electron description to a pseudo-potential one through appropriate groupings of electrons with effective $g^{-1}_0$ values, for example, dividing electrons up into core and valence shells in order to speed computations. It is also apparent that if one keeps rigorously only pairs of electrons, and solves the zero temperature case by expanding the diffusion equations (\ref{diff4}) in basis sets of the eigenfunctions of the spatial operators of (\ref{diff4}), as in reference \onlinecite{Thompson2019} appendix B, then one will regain KS-DFT. Thus depending on the grouping of electrons, the present approach reproduces both KS-DFT and OF-DFT.

\begin{figure}
\begin{tabular}{cc}
\includegraphics[width=0.45\textwidth]{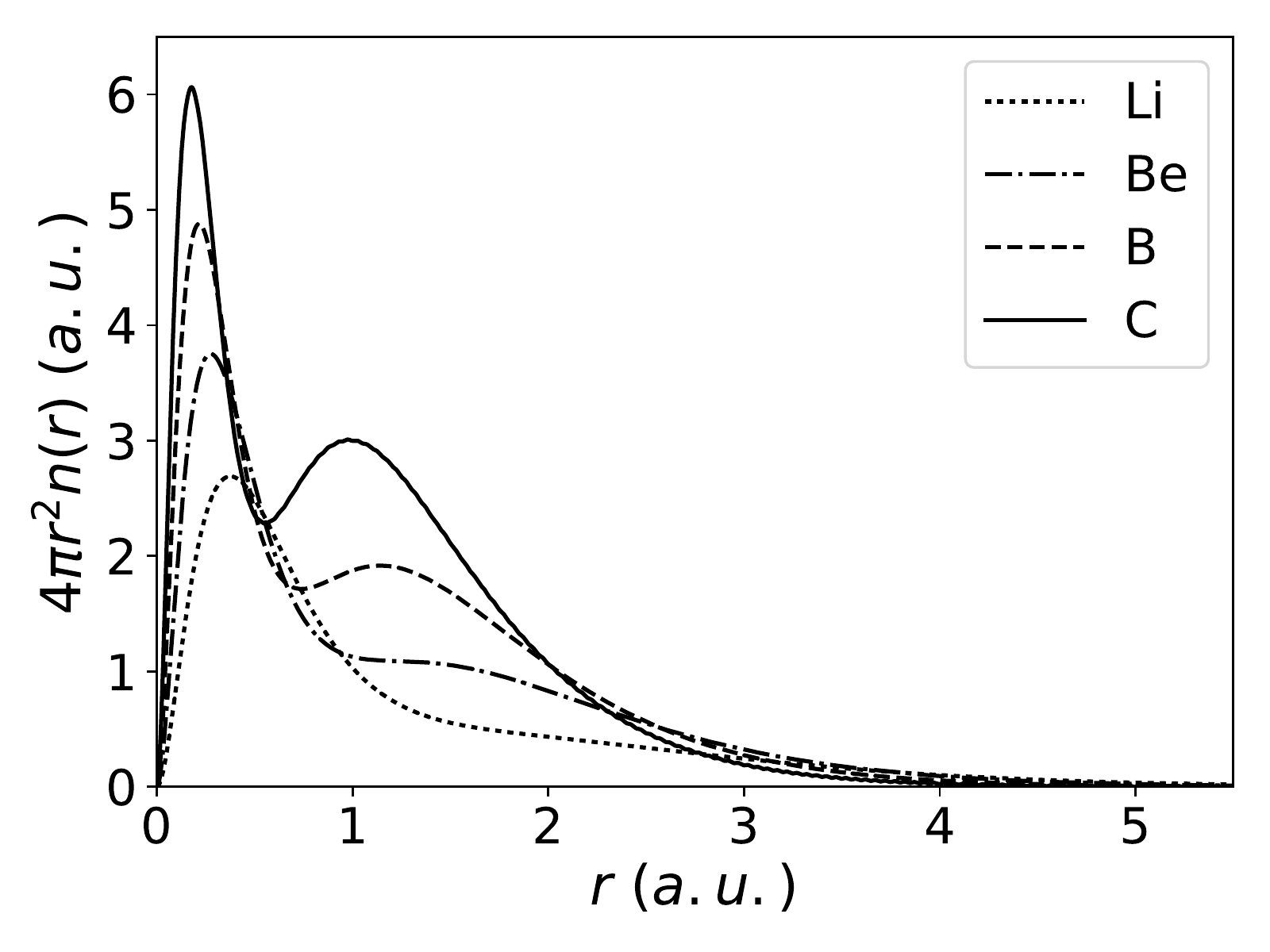} & \includegraphics[width=0.45\textwidth]{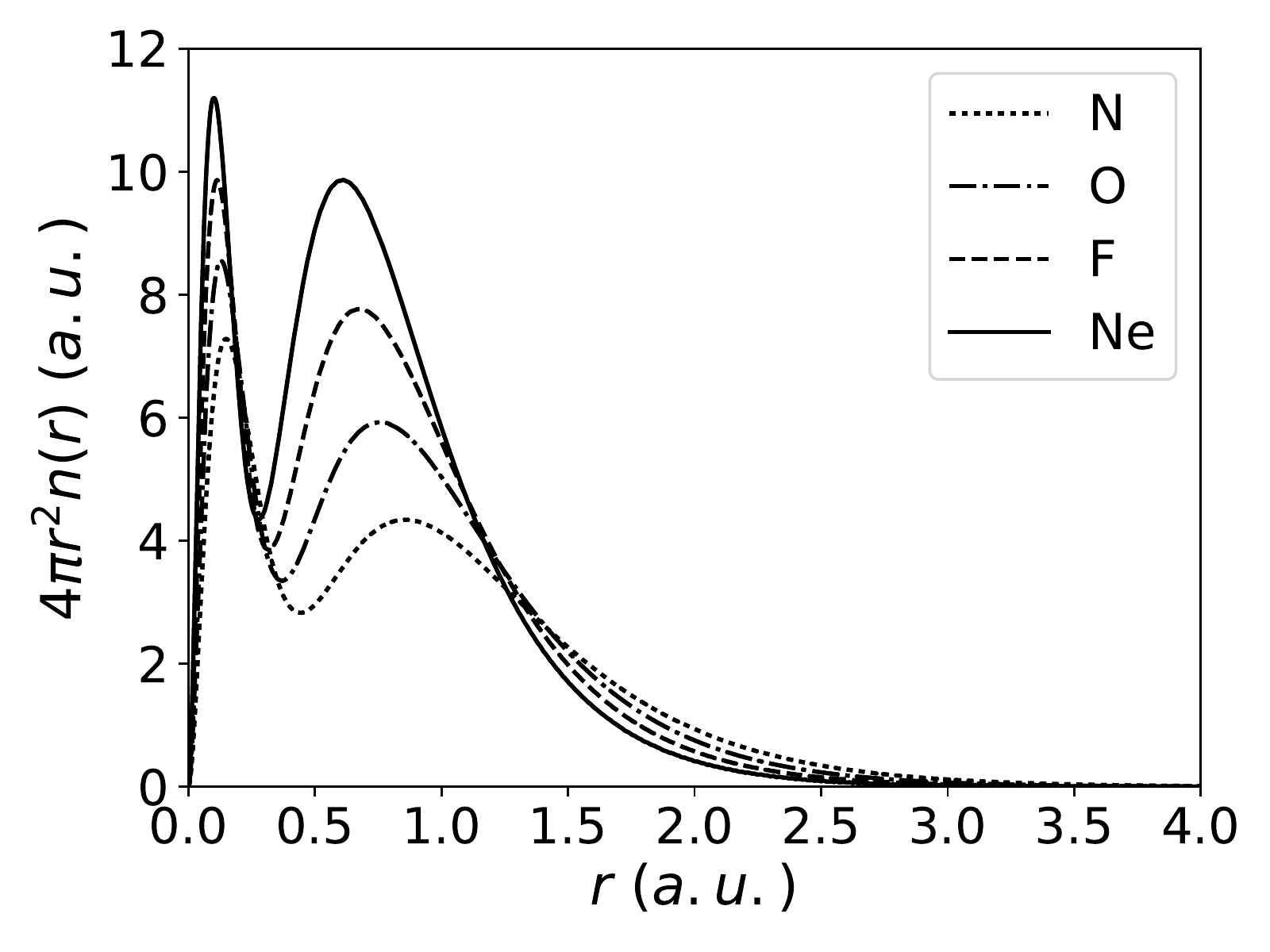} \\
(a) & (b) \\
\includegraphics[width=0.45\textwidth]{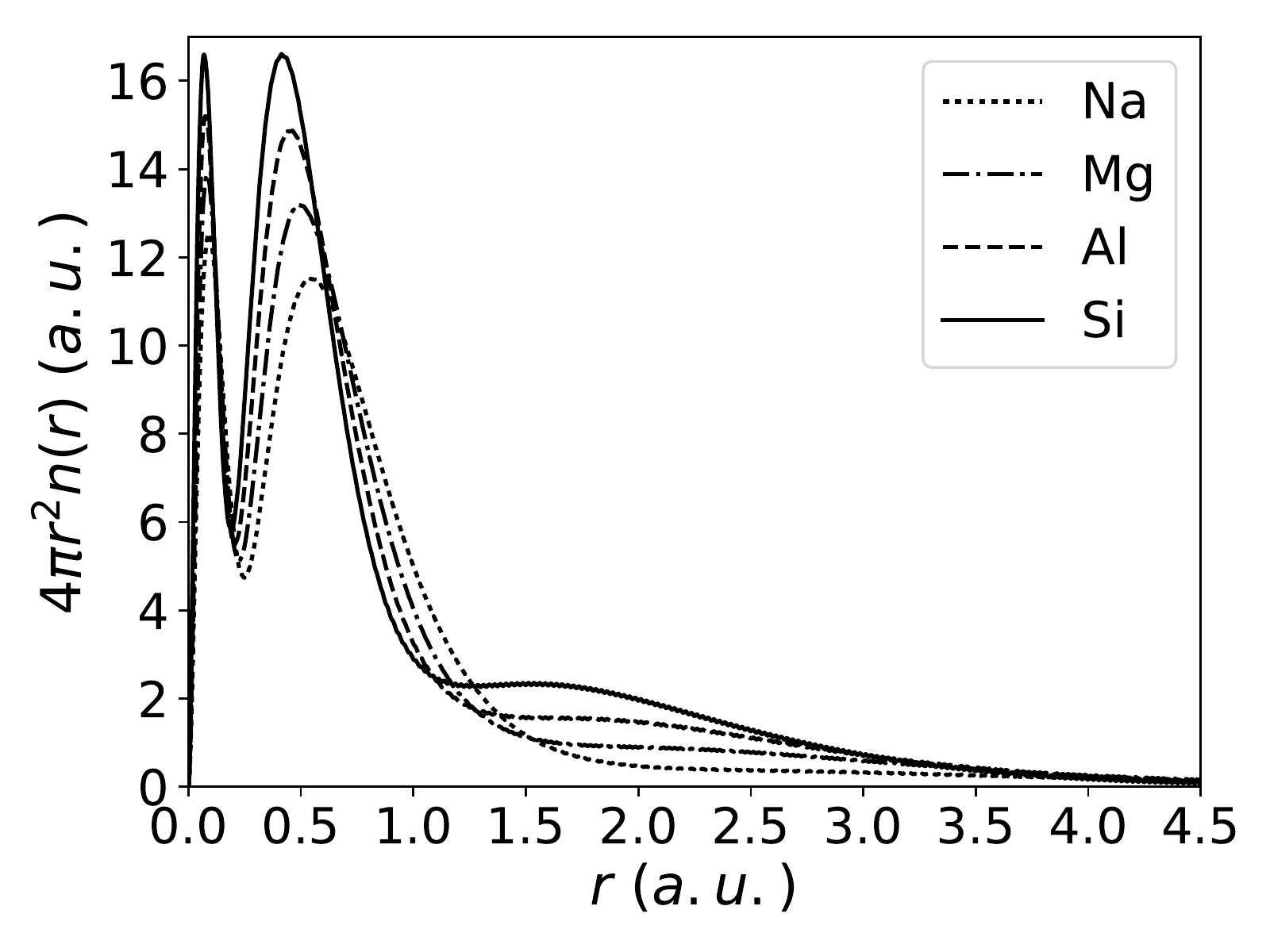} & \includegraphics[width=0.45\textwidth]{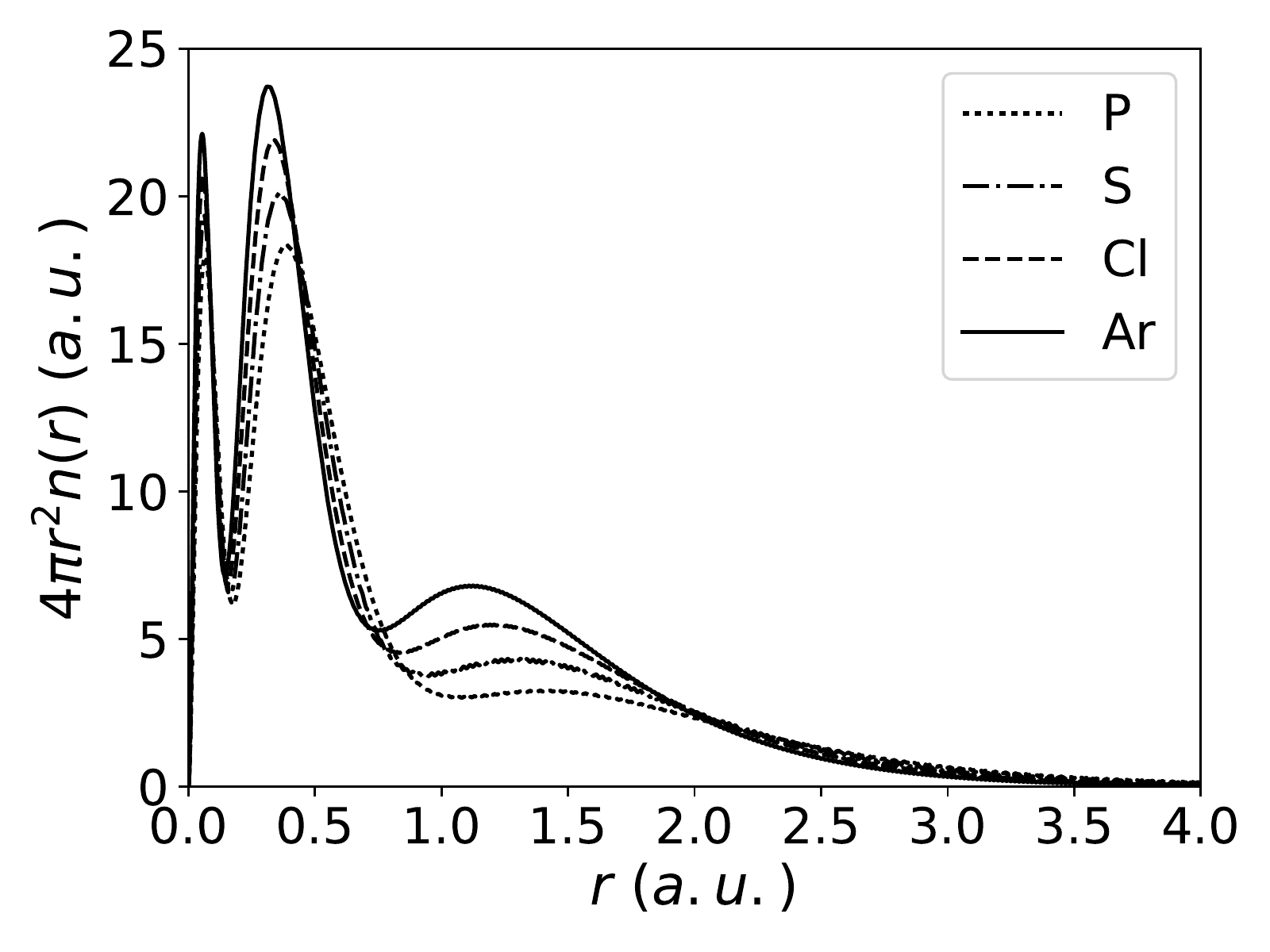} \\
(c) & (d) 
\end{tabular}
\caption{Plot of radial electron densities as a function of radius, in atomic units, for (a) lithium, beryllium, boron and carbon atoms; (b) nitrogen, oxygen, fluorine and neon atoms; (c) sodium, magnesium, aluminum and silicon atoms; (d) phosphorus, sulphur, chlorine and argon atoms.}
\label{fig:densities}
\end{figure}

Radial electron densities are shown in figure \ref{fig:densities}. 
All densities show correct shell structure and proper magnitudes. The densities and corresponding binding energies from table \ref{tab:elemFE} can be compared with the predictions of the simple shell-structured-based (SSB) Pauli potential of Finzel \cite{Finzel2015}, which was used in reference \onlinecite{Thompson2019}. The electron densities are observed to be qualitatively the same, with some small quantitative differences. The binding energies here, as mentioned, are quite good, in contrast to those that can be calculated using the Finzel SSB Pauli potential, which are very far from NIST values. The values of Finzel can be improved somewhat by dropping the LDAX term and incorporating instead the Fermi-Amaldi correction to the electron-electron self-interaction. This implies the exclusion effect is double counted when an explicit Pauli potential is used with the LDAX. An exception to this is helium, for which the binding energy becomes worse because there is no Pauli potential and the Fermi-Amaldi expression becomes poor. Aside from helium, to achieve these improvements for binding energy using the SSB Pauli potential, one must inconsistently add an arbitrary term involving the Pauli potential $w_P({\bf r})$ to the free energy since, as shown in reference \onlinecite{Thompson2019}, the SSB Pauli potential behaves mathematically as an external potential. This is because the step functions on which it is based are not calculated from the electron density, but are input ``by hand'' based on other arguments. The contour excluded volume method of incorporating the Pauli potential in this paper does not suffer this fault and produces better binding energies than the SSB potential (with the exception of helium) in all situations. It is also flexible, since it is not based on a spherical step function, but is self-consistently derived from the electron densities, and so it can be applied to molecules and solid-state materials. Given the Fourier spectral method of solution, it should be especially well suited to the latter, although this is beyond the scope of the present work. 

The results shown in table \ref{tab:elemFE} can also be compared to other orbital-free methods based on Thomas-Fermi-Dirac-Weizsacker (TFDW) approaches --- see reference \cite{Dreizler1990} for example. The present ground state binding energies are better than some TFDW functionals and worse than others, but it must be emphasized that the TFDW methods do not produce correct shell-structures as does the current method and the SSB method of Finzel. Also, as mentioned, the Fermi-Amaldi self-interaction term used here is very basic and no correlations have been included. There is thus scope for improvement in the binding energies. 

In principle, the computational efficiency of the present method should range from about the same as OF-DFT to about the same as KS-DFT. This would depend on how many explicit “shells” are used in the computation. A single shell (no excluded volume between any electron ``contours'') would yield a single diffusion equation to solve, whereas enforcing the Pauli principle completely would give $N/2$ diffusion equations. These $N/2$ equations are independent and so could be trivially solved in parallel. The ability to smoothly go from OF-DFT to KS-DFT could allow for some beneficial approximation schemes. A direct benchmarking between the present code and standard OF-DFT or KS-DFT codes is beyond the scope of this work because available codes have been optimized over many years whereas no significant attempt at numerical efficiency has been incorporated into the preliminary code used in this work.

\section{Discussion}

The above results and theory are achievable using only classical concepts in a four dimensional thermal-space. Two postulates are needed for the formalism, by which in principle all static and non-relativistic quantum mechanical results can be obtained.

The first postulate is from reference \onlinecite{Thompson2019} and is that quantum particles are classically modelled as Gaussian threads in a four dimensional thermal-space. From the application of classical statistical mechanics through the SCFT formalism, ensemble average predictions give the same results as static quantum DFT, which in turn gives the same results as static, non-relativistic quantum mechanics, as proven by the theorems of DFT \cite{Hohenberg1964, vonBarth2004, Becke2014, Jones2015}. Note that in the derivation of reference \onlinecite{Thompson2019}, DFT theorems are not needed. This first postulate can be shown to be equivalent to the Heisenberg uncertainty principle. 

One way of expressing the uncertainty principle is through the commutation relation between position and momentum operators in standard quantum mechanics. Position and momentum are conjugate quantities in that one is the Fourier transform of the other. In the position representation, if the eigenfunctions of the position operator are Dirac delta functions, then the eigenfunctions of the momentum operator must be $\sim \exp (\frac{i}{\hbar} {\bf p}\cdot{\bf r})$ to be conjugate to the position eigenfunctions. If the uncertainty relation does \emph{not} hold, then the momentum eigenfunctions will not be $\exp (\frac{i}{\hbar} {\bf p}\cdot{\bf r})$ with respect to the Dirac delta position eigenfunctions. The textbook explanation of McQuarrie \cite{McQuarrie2000} used in the derivation of the diffusion equation description of quantum particles in reference \onlinecite{Thompson2019} will no longer hold, and quantum particles will \emph{not} be described by contours in a 4D thermal-space. In other words, the Fourier transform of the governing diffusion equation, equation (A14) of reference \onlinecite{Thompson2019}, gives the relationship between momentum and position at all temperatures assuming position and momentum are conjugate quantities (reference \onlinecite{Thompson2019} equation (A15) \footnote{Equation (A15) of reference \onlinecite{Thompson2019} is missing a factor of complex $i$ in the argument of the exponential.}), which is the uncertainty relation. More directly, for position and momentum to commute, one must have $\hbar \rightarrow 0$, that is, the classical limit. This limit of the diffusion equation gives just classical statistical mechanics for point-like particles as shown in appendix C of reference \onlinecite{Thompson2019}. Thus the polymeric description of quantum particles governed by a diffusion equation is related to the validity of the uncertainty principle. The postulate of quantum particles being Gaussian threads in a classical 4D thermal-space is equivalent to postulating the Heisenberg uncertainty principle in a quantum 3D space. 

The second postulate is that the Gaussian threads have excluded volume in 4D thermal-space with respect to occupancy beyond two threads (in the case of electrons). The results given in this paper demonstrate that this postulate gives the correct shell structure for atoms expected from the Pauli exclusion principle. Thus postulating excluded volume Gaussian threads in classical 4D thermal-space is equivalent to postulating the Pauli exclusion principle in quantum 3D space.

These two 4D postulates, with classical statistical mechanics, are equivalent to static and non-relativistic quantum mechanics. At least one more postulate would be required to describe the dynamics of systems, but that is beyond the scope of this work.

\section{Conclusions}

The Pauli exclusion principle is a fundamental feature of nature, underpinning classical excluded volume. In this work, a return to an excluded volume postulate, but in a 4D thermal-space, has been shown to quantitatively reproduce atomic shell structure. This postulate, together with a postulate stating that quantum particles are polymer-like Gaussian threads in the thermal-space, is sufficient to describe static, non-relativistic quantum mechanics. By taking this classical statistical mechanics thermal-space perspective, the measurement problem is avoided, since there are no wave functions to collapse. Instead, wave-particle duality arises from the non-locality of polymer-like quantum particles, as discussed in reference \onlinecite{Thompson2019}. Since this polymer-like non-locality is also related to the Heisenberg uncertainty principle, as discussed in the previous section, it is seen that the uncertainty principle and wave-particle duality are connected. This is consistent with the ideas of Coles et al. \cite{Coles2014}, who showed that wave-particle duality is equivalent to entropic uncertainty. In the SCFT formalism, the solution of the diffusion equation describing quantum particle contours gives the conformational entropy in polymer molecules. That is, as mentioned in reference \onlinecite{Thompson2019}, the quantum kinetic entropy term is equivalent to polymer conformational entropy, connecting entropy to the uncertainty principle as suggested by Coles et al. \cite{Coles2014}. This idea may be worth exploring more deeply in the future.

The representation of the Pauli exclusion principle through classical excluded volume is related to the ideas of Hayakawa and Hong, who showed that Fermi statistics arise classically in theoretical two dimensional granular systems due to excluded volume \cite{Hayakawa1997}. The broader approach of obtaining aspects of quantum mechanics through classical statistical mechanics connects to the ideas of other groups \cite{Budiyono2017, Huang2004}. The use of statistical mechanics to describe quantum mechanics requires a thermal interpretation of quantum physics, which connects the SCFT approach to ideas of Neumaier \cite{Neumaier2019}. It could be useful in the future to explore possible relationships between SCFT and these other methods. 

Despite the relationships with other work, there are unaddressed issues for a 4D thermal Gaussian thread interpretation of quantum mechanics. In reference \onlinecite{Thompson2019}, non-locality of Gaussian threads was used to speculate on the double slit experiment, that is, to explain wave-particle duality. In fact, that speculative connection was only partially accomplished due to the equilibrium nature of the SCFT formalism. The arguments of reference \onlinecite{Thompson2019} justified Huygen's principle, showing that SCFT results will differ from classical predictions for the double-slit experiment, but a dynamic description including the de Broglie wavelength will be needed to predict a full interference pattern. Nonetheless, the static formulation with the versatile Pauli potential given in this work could be applied to any number of molecular and solid state systems. The periodic feature of solid state materials means the spectral solution method could be very practical. For molecular systems, one might consider using basis functions based on Gaussians, which is standard in computational chemistry. For both system types, the use of effective inverse density of states parameters $g^{-1}_0$ in the context of pseudo-potentials could reduce the computational burden significantly, in keeping with the spirit of OF-DFT, and allow practical DFT calculations for many applications.

\section*{Acknowledgements}

The author thanks V. V. Ginzburg for commenting on the manuscript prior to publication. This research was financially supported by the Natural Sciences and Engineering Research Council of Canada (NSERC).

\appendix

\section{Uniform Limits}

Two uniform, high electron number, expressions for the Pauli energy density are given in this appendix. First, the mean field expression (\ref{UP1}) for the Pauli energy can be written in the uniform limit easily as
\begin{equation}
U_P = \frac{V}{2} \sum_{ij} g^{-1}_{ij} n_{i0}n_{j0}   \label{UP3}
\end{equation}
where $n_{i0}$ is the uniform density for the pair of particles with density $n_i({\bf r},\beta)$ and sums are over all pairs. Density $\beta$ dependence will be suppressed for clarity. Dividing through by $V$ and writing the uniform densities as $n_{i0} = N_i/V$, where $N_i$ is the number of electrons of each pair, 0, 1 or 2, gives
\begin{equation}
\frac{U_P}{V} = \frac{1}{2V^2} \sum_{ij} g^{-1}_{ij} N_{i}N_{j}   . \label{UP4}
\end{equation}
Assuming a large number of total electrons, it is safe to choose all $N_i = 2$ without loss of generality. This gives
\begin{eqnarray}
\frac{U_P}{V} &=& \frac{4}{2V^2} \sum_{ij} g^{-1}_{ij} \nonumber \\
&=& \frac{2g^{-1}_0}{V^2} \left(\frac{N^2}{4} - \frac{N}{2} \right)  \label{UP5}
\end{eqnarray}
where $N$ is the total number of electrons. Equation (\ref{g0inv}) is used to get (\ref{UP5}), as is the fact that the sums run over the number of electron pairs, that is, up to $N/2$. For large numbers of electrons, the second term in (\ref{UP5}) can be dropped, giving 
\begin{equation}
\frac{U_P}{V} = \frac{g^{-1}_0}{2} n_0^2  \label{UP6}
\end{equation}
which is equation (\ref{TF2}).

Obviously, equation (\ref{UP6}) is not the correct Thomas-Fermi limit given by (\ref{TF}), which might cause concern as to whether the model of polymeric-type excluded volume is consistent with the Pauli exclusion principle. To test this, one can avoid the mean field approximation and use polymer scaling theory to see if the uniform, high electron number situation for polymer excluded volume scales correctly. The following argument is an adaptation of the ideas of de Gennes \cite{deGennes1979}. 

For high electron number density, and ignoring spin, which will only add an overall factor of two, each 4D electron polymer-contour will be confined to a four-dimensional hyper-cylinder due to the presence of other electron polymer-contours surrounding it. This is similar to the reptation concept in entangled polymer dynamics \cite{deGennes1979}. The energy of a single electron-polymer confined to a tube scales as \cite{deGennes1979}
\begin{equation}
\frac{U_P}{N} \sim D^{-2}   \label{scale1}
\end{equation}
where $N$ is the number of electrons in the volume $V$ and $D$ is the cross-section of the hyper-cylinder. In the 4D thermal-space, the cross-section is the volume per single electron, thus $D \sim n_0^{-1/3}$ where $n_0$ is the uniform density of electrons. Thus (\ref{scale1}) becomes 
\begin{equation}
\frac{U_P}{N} \sim n_0^{2/3}  .  \label{scale2}
\end{equation}
This is for a single electron. To get the total energy per volume, one multiplies through by $n_0 = N/V$ giving
\begin{equation}
\frac{U_P}{V} \sim n_0^{5/3}    \label{scale3}
\end{equation}
which is equation (\ref{TF3}).

\bibliography{DFTbibliography5}

\end{document}